\documentclass{aastex}
\usepackage{natbib}
\usepackage{emulateapj5}

\shorttitle{The galactic bulge formation}
\shortauthors{Nakasato \& Nomoto}

\newcommand{\SM}{$M_{\odot}$}

\begin{document}

\submitted{Appear in the Astrophysical Journal}

\title{3-D Simulations of the Chemical and Dynamical Evolution of the Galactic Bulge}

\author{Naohito Nakasato\altaffilmark{1,2} and Ken'ichi Nomoto\altaffilmark{1}}

\altaffiltext{1}{Department of Astronomy, School of Science,
University of Tokyo, Bunkyo-ku, Tokyo 113-0033;
nakasato@astron.s.u-tokyo.ac.jp, nomoto@astron.s.u-tokyo.ac.jp}

\altaffiltext{2}{Astronomisches Rechen-Institut, Heidelberg, 69120}


\begin{abstract}
A three-dimensional hydrodynamical N-body model for the formation of
the Galaxy is presented with special attention to the formation
of the bulge component.
Since all previous numerical models for the Galaxy formation do not
have a proper treatment of the chemical evolution and/or 
sufficient spatial resolutions, we have constructed a detailed model of
the chemical and dynamical evolution of the Galaxy using our GRAPE-SPH code.
Our SPH code includes various physical processes related to
the formation of stellar systems.
Starting with cosmologically motivated initial conditions, 
we obtain a qualitatively similar stellar system to the Galaxy.
Then we analyze the chemical and kinematic properties
of the bulge stars in our model and find qualitative agreement
with observational data. 
The early evolution of our model has revealed that
most bulge stars form during the sub-galactic merger
(merger component of the bulge stars).
Because of the strong star burst induced by the merger, the metallicity
distribution function of such stars becomes as wide as observed.
We find that another group of the bulge stars forms later
in the inner region of the disk (non-merger component of the bulge stars).
Because of the difference in the formation epoch,
the main source of iron for this group of stars
is different from the merger component.
Iron in the merger and non-merger components
comes mainly from Type II and Type Ia supernovae, respectively.
Since a Type Ia supernova ejects $\sim$ 10 times more iron than
a Type II supernova, [Fe/H] of the non-merger component
tends to be higher than that of the merger component, 
which widens the metallicity distribution function.
From these results, we suggest that the Galactic bulge
consists of two chemically different components;
one has formed quickly through the sub-galactic clump merger
in the proto-galaxy and the other has formed gradually in the inner disk.
\end{abstract}


\section{Introduction}
The formation of bulges, 
which are the high density central component of spiral galaxies,
is a key process in the formation of spiral galaxies.
There are several scenarios for the formation process of the bulges
\citep[e.g.,][]{Bouwens_1999}.
Such scenarios can be divided into the following categories;
(1) a primordial collapse \citep[e.g.,][]{Eggen_1962}, 
(2) merging of sub-clumps in a proto-galactic cloud
\citep[e.g.,][]{Baugh_1996}, and
(3) a secular evolution of a stellar disk \citep[e.g.,][]{Norman_1996}.
Previously, the surface brightness of the bulges has been
recognized to have the $R^{1/4}$ profile as in ellipticals,
where $R$ denotes the distance from the center
\citep[e.g.,][]{Whitford_1978, Rich_1988}.
However, recent observations have revealed the presence of a class
of bulges that are well fitted by the exponential profile
\citep[e.g.,][]{Andredakis_1995}, 
i.e., there exist two classes of bulges according to the surface brightness.
Each formation scenario has advantage and disadvantage to explain
the diversity of bulges \citep{Wyse_1999}.

For the formation of the Galactic bulge, 
\citet{Matteucci_Brocato_1990} and \citet{Matteucci_1999} have 
constructed one-zone chemical evolution models as follows:
(1) They assumed that the Galactic bulge was formed by a single star-burst.
In their best fit model, the formation time scale is $\sim$ 0.1 Gyr.
(2) One of the predictions of their calculations is that
the element ratio such as [O/Fe] remains high even for metal-rich stars
([Fe/H] $>$ 0) because the formation time scale of the bulge
is shorter than the lifetime of a Type Ia supernova progenitor.
However, the chemical properties predicted by the one-zone model depend
strongly on their {\it assumption} of the dynamics, i.e.,
the formation time-scale of the bulge.
To constrain the formation scenarios, we need to calculate
the chemical and dynamical evolution together and compare the model
predictions with both the chemical and kinematic properties of the bulge.

Here, we present our first results of the three-dimensional Galactic bulge formation model.
We model the formation of the Galaxy by means of Smoothed Particle 
Hydrodynamics (SPH) method \citep{Lucy_1977, Gingold_Monaghan_1977}
that incorporates a star formation algorithm
and a self-consistent treatment of the chemical enrichment history of gas.
The three-dimensional models for the formation of disk galaxies 
have been investigated by many authors
\citep[e.g.,][]{Katz_1992, Steinmetz_Muller_1995, Friedli_Benz_1995, Berczik_1999}
and they have succeeded in many respects.
We follow the method and models by the previous authors
but use much larger number of particles to study
the detailed formation and evolution processes of the Galaxy \citep{Nakasato_thesis}.

In Cold Dark Matter (CDM) cosmology, the early stage of the formation of
galaxies involves the progressive merging of sub-galactic clumps 
\citep[e.g.,][]{Baugh_1996}.
To properly model the dynamics of the merger history, 
we adopt the three-dimensional SPH method.
The chemodynamical approach by \cite{Samland_1997}, 
which is a mesh based and two-dimensional method, 
is not suitable to model the early merging history of the proto-galaxy.

Our SPH code uses the GRAPE \citep{Sugimoto_1990} and 
the Remote-GRAPE system \citep{Nakasato_1997} to compute the gravitational
interaction between particles (GRAPE-SPH code).
The adoption of the Remote-GRAPE system doubles the performance of our code
since we can make the SPH calculation and N-body summation in parallel
\citep{Nakasato_1997}.
This enhancement of performance enables us to use seven times 
larger number of particles than \citet{Katz_1992},
\citet{Steinmetz_Muller_1995}, and \citet{Raiteri_1996}
so that the spatial resolution of our model is almost two times higher
than their models.
As a result, we can investigate the early merging history
in much more detail during the Galaxy formation.
We note that other authors \citep[e.g.,][]{Navarro_Steinmetz_2000, SommarLarsen_2001}
have published the comparable or even higher resolution SPH galaxy formation 
models than the present work; however, these authors have mainly concentrated
on the kinematic properties of the models.
In contrast, in the present work, we have made 
attention to both the chemical and kinematic properties.

Our study is the first attempt to model the formation and
chemical evolution of the Galactic bulge using the self-consistent
three-dimensional SPH method.
Using the chemical and dynamical SPH code, we simulate
the evolution of the Galaxy starting from a plausible cosmological
initial condition.
The obtained model well reproduces the important chemical and kinematic
properties of the Galaxy such as the formation of the bulge,
disk and halo as seen in Figure \ref{3comp}.
By analyzing the result of this model, 
we construct a plausible scenario for the Galactic bulge formation.

The plan for the present paper is as follows:
In section 2, we briefly describe our SPH method.
Section 3 describes the procedure to construct initial conditions.
We present a summary of results obtained with our model in section 4.
By analyzing the result of our Galaxy model, 
we describe the formation history of the bulge component in section 5, 
and compare the chemical and kinematic properties of the bulge
with recent observational results.
Finally, section 6 is devoted to conclusions.

\section{The method}

We adopt the SPH method to construct
self-consistent three-dimensional dynamical and chemical
models for the formation of stellar systems.
The SPH method has been applied to many astrophysical problems.
Because of its Lagrangian nature, it is suitable to systems
that have large density contrasts,
e.g., the formation of galaxies
\citep[e.g.,][]{Evrard_1988, Hernquist_Katz_1989, Katz_1992, Steinmetz_Muller_1994},
the evolution of galaxies \citep[e.g.,][]{Friedli_Benz_1995, Patsis_2000},
the cosmological simulations \citep[e.g.,][]{Navarro_White_1994, Yoshikawa_2000},
and a cloud-cloud collision \citep[e.g.,][]{Lattanzio_1985, Habe_Ohta_1992}.
Various codes have been developed to combine SPH and N-body systems.
In these codes, gravitational forces are calculated
with various methods such as direct summations,
Particle-Particle/Particle-Mesh methods \citep[e.g.,][]{Evrard_1988},
Tree methods \citep[e.g.,][]{Hernquist_Katz_1989, Benz_1990},
and the method to use the special purpose computer GRAPE
\citep[e.g.,][]{Umemura_etal_1993, Steinmetz_1996}.

To simulate the formation of a stellar system from gas,
we use our GRAPE-SPH code using the Remote-GRAPE library \citep{Nakasato_1997}.
The SPH formulation that we use is the same as \citet{Navarro_White_1993}.
We use a spatially variable smoothing length and integrate equations
of motion with a second order Runge-Kutta method as described
in \citet{Navarro_White_1993}.
To simulate the formation of stellar systems from gas,
our GRAPE-SPH code includes various physical processes related
to the formation of stellar systems, e.g., cooling, star formation, 
and feedback by stars as follows.

\subsection{Cooing and star formation}
We adopt the metallicity dependent cooling functions, 
which are computed by MAPPINGS III package by R.S. Sutherland
\citep[MAPPINGS III is the successor of MAPPINGS II described in][]{Sutherland_Dopita_1993}, 
to solve the energy equation of gas particles
(see Figure 1 of \citet{Nakasato_2000} for the actually used cooling functions)

For the star formation algorithm, we adopt the star formation recipe
as used in the usual SPH codes \citep{Katz_1992}.
Hereafter, ``STAR'' means ``star particle'', 
which is not an individual star but an association of many stars.
In this recipe, we convert a fraction of a gas particle
into a STAR when four conditions are satisfied.
The first three conditions are the {\it physical} conditions expressed as
\begin{equation}
(\nabla \cdot \mbox{\boldmath{$v$}})_i < 0,
\label{sf_1}
\end{equation}
\begin{equation}
t_{\rm cool} < t_{\rm d},
\label{sf_2}
\end{equation}
\begin{equation}
t_{\rm d} < t_{\rm sound}, 
\end{equation}
where $t_{\rm cool}$, $t_{\rm d}$ , and $t_{\rm sound}$
are the cooling time, dynamical time, and sound crossing time
of the gas particle, respectively
(see \cite{Nakasato_2000} for the detailed time scale definition).

\begin{figure}
\plotone{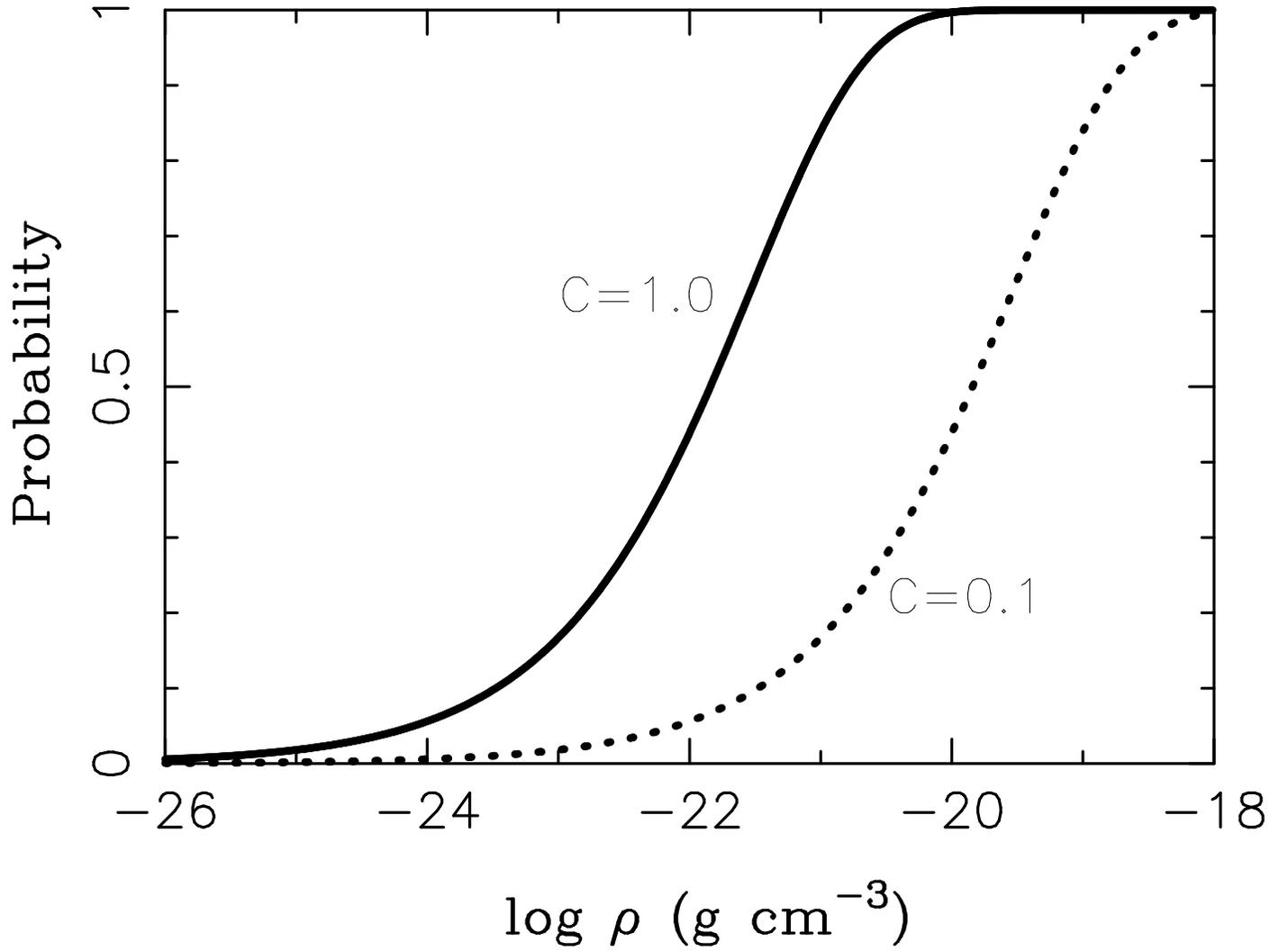}
\caption{
The star formation probability ($P$) as a function of density of gas.
The solid and dashed lines correspond to $C =$ 1.0 and $C =$ 0.1, 
respectively.
\label{sf_prob}
}
\end{figure}

During each star formation interval ($\delta t = 2$ Myr in the present work), 
we check whether the three conditions are satisfied for each gas particle.
If the gas particle satisfies these three conditions, 
we then check the forth {\it probability} condition.
Assuming star formation process is a Poisson process
\citep{Katz_1992}, 
we can write the probability for star formation within $\delta t$ as
\begin{equation}
P = 1.0 - {\rm exp}(-\delta t/t_{\rm starform}), 
\end{equation}
where $t_{\rm starform}$ is the local star formation time scale.
In our SPH code, we define $t_{\rm starform} = t_{\rm d} / C$, 
where $C$ is a parameter to control our star formation recipe.
If $P$ is larger than a random number $(0 - 1)$, 
we create a new STAR from a fraction of the gas particle
(star formation occurs!).
Introducing this {\it probability} condition, we can effectively 
control the threshold density for star formation by changing the parameter $C$.
We compute $P$ as a function of the density ($\rho$)
for different values of $C$ as shown in Figure \ref{sf_prob}.
Here, we assume $\delta t =$ 2 Myr.
Note that the timestep used in the time integration
is always smaller than $\delta t$.
From Figure \ref{sf_prob}, we can see effects of $C$ as follows;
larger (smaller) $C$ means lower (higher) threshold density.

\begin{figure}
\plotone{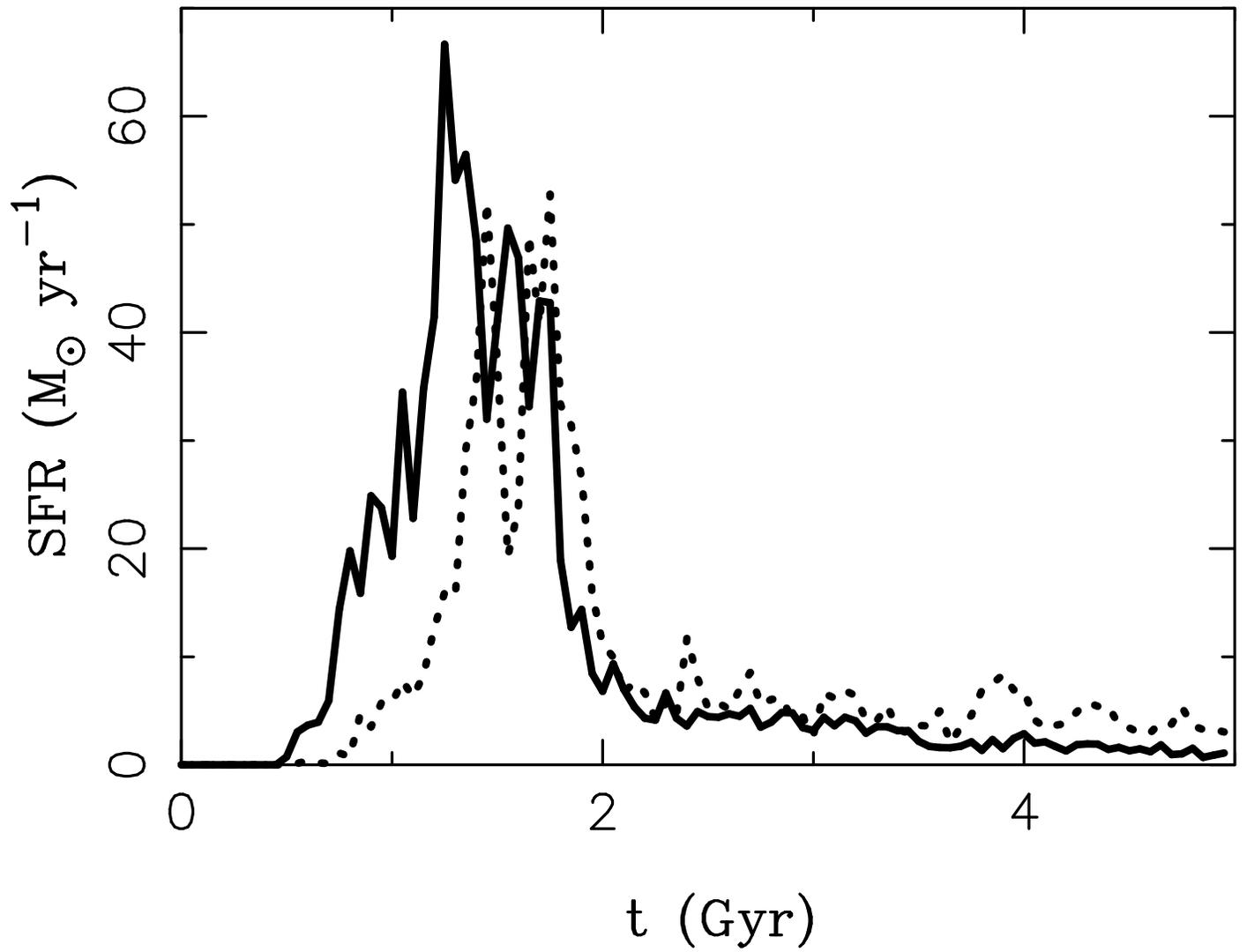}
\caption{
The star formation rate (SFR) as a function of time for 
galaxies with $C =$ 0.1 (the dotted line)
and $C =$ 1.0 (the solid line).
\label{prev_sfr}
}
\end{figure}

To demonstrate the dependence on the star formation parameter $C$, 
we evolve a spherical region of 10$^{12}$ \SM, 
which consists of 10 \% gas and 90 \% dark matter,
for various $C$.
Initially, we set up the sphere in rigid rotation
and outward expansion.
And the angular speed of rotation corresponds to
the spin parameter $\lambda \sim$ 0.05, 
where $\lambda = \frac{L |E|^{1/2}}{G M^{5/2}}$,
with $L$ being the angular momentum, $E$ the total energy, 
M the mass, and $G$ the gravitational constant \citep{Padmanabhan_1993}.
This initial condition is the same as used in \citet{Katz_1992}
and \citet{Steinmetz_Muller_1995}.
We evolve the sphere for three different star formation parameters
$C =$ 0.1, 0.5 and 1.0.
In Figure \ref{prev_sfr}, we present the star formation history
for $C =$ 0.1 and $C =$ 1.0.
For $C =$ 1.0, the star formation rates (SFR) increase more rapidly than $C =$ 0.1
since the threshold density for star formation is lower.
These STARs which formed early (i.e., very old STARs)
in the $C =$ 1.0 model have large vertical velocities.
If we compare the projected edge-on surface density of STARs, 
the $C =$ 1.0 model clearly shows more extended distribution then the $C =$ 0.1 model.
In Table \ref{dyn}, some kinematic properties at $t =$ 5 Gyr
are compared for different $C$.
There is a clear correlation between $C$ and 
the ratio between the X and Z axis velocity-dispersion (or the spin parameter).
From these results, we conclude that the $C =$ 0.1 and $C =$ 1.0
galaxies are similar to late type and early type galaxies, respectively. 
To summarize, if we use this {\it probability} condition,
various values of $C$ lead to various global star formation histories.
Hereafter, we set $C =$ 0.1 throughout the present paper.

The mass of a newly formed STAR ($M_{\rm star}$) is calculated as
\begin{equation}
M_{\rm star} = \left[ 1 - \frac{1}{1 + 0.5 \frac{\delta t}{t_{\rm starform}}}
\right]
\pi h^3 \rho, 
\end{equation}
where $h$ and $\rho$ are the smoothing length and the density
of the gas particle, respectively \citep{Nakasato_2000}.
In the present calculation, the typical mass of STARs is $\sim 10^6 - 10^7$ \SM.

\subsection{Feedback from stars}
Formed stars eject energy and mass
as stellar winds and supernova explosions (feedback from stars).
As a result, stars heat up, accelerate, and enrich
the circumstellar and interstellar medium.
With current computing resources, it is not yet feasible
to accurately include the energy, momentum, and mass 
feedback from stars in our SPH code.
This is because the length resolution of the present calculation
($\sim 0.5$ kpc) is much longer than a typical size of
supernova remnants ($< 100$ pc).
Thus, in the present paper, we adopt a phenomenological method
as adopted by several authors \citep[e.g.,][]{Katz_1992}
to mimic real feedback processes;
i.e., we distribute the energy and mass ejected
by a STAR to the neighbor gas particles.
In our SPH code, we incorporate the effect of stellar winds and Type II and Type Ia
supernova explosions as the energy and mass feedback processes.
We compute the mass and energy ejection rates based on a simple model
explained below.
In this paper, we distribute the feedback energy as purely thermal energy.
This treatment is a simple and a zero-th order approximation to the real nature
and other authors have tried to refine the treatment of feedback
\citep[e.g.,][]{Thacker_2000, Springel_2002}.
Also in this paper, we follow the chemical evolution
of total metal (Z), iron (Fe) and oxygen (O).
The chemical evolution is coupled with the thermal and dynamical
evolution, i.e., the ejected metal affects the evolution of gas
through the metallicity dependent cooling rates.
To summarize, our GRAPE-SPH code is a most up-to-date SPH implementation
for the formation of stellar systems and used to successfully model
the formation of globular clusters \citep{Nakasato_2000}.

\subsection{Energy ejection}
The energy ejection rate per STAR is given as
\begin{equation}
E_{\rm eject} = e_{\rm SW} R_{\rm SW} + e_{\rm SNII} R_{\rm SNII}
	 + e_{\rm SNIa} R_{\rm SNIa},
\end{equation}
where $e_{\rm SW}$ is the total ejected energy by stellar winds during
the stellar life time and $e_{\rm SNII}$ and $e_{\rm SNIa}$ are
the energies ejected by a Type II and Ia supernova explosion, respectively.
The rate $R_{\rm SW}$ is the number of stars that expel
their envelopes at the current epoch per unit time
and $R_{\rm SNII}$ and $R_{\rm SNIa}$ are
the rates of Type II and Type Ia supernovae, respectively.
We define $R_{\rm SW}$ and $R_{\rm SNII}$ as follows
\begin{equation}
R_{\rm SW} = \frac{\displaystyle{\int^{M_{\rm up}}_{M_{\rm ms}} \phi(m) dm}}
{\displaystyle{\tau(M_{\rm ms})}}
\label{R_SW}
\end{equation}
\begin{equation}
R_{\rm SNII} = \frac{\displaystyle{\int^{M_{\rm ma}}_{M_{\rm ms}} \phi(m) dm}}
 {\displaystyle{\tau(M_{\rm ms}) - \tau(M_{\rm ma})}}, 
\label{R_SNII}
\end{equation}
where $\phi(m)$ is the initial mass function (IMF), namely
$\phi(m) dm$ gives the number of stars in the mass range of $(m, m+dm)$,
and $\tau(m)$ is the stellar lifetime as a function of stellar mass
\citep{David_1990}.
In the present study, we assume the power-law type IMF as
\begin{equation}
\phi(m) = A m^{-2.35}, 
\end{equation}
where $A$ is a constant.
For the upper and lower mass limits in the IMF, 
$M_{\rm up} =$ 120 \SM~ and $M_{\rm lo} =$ 0.05 \SM~ are assumed.
In Eqs. (\ref{R_SW}) and (\ref{R_SNII}),
$M_{\rm ma}$ ($=$ 50 \SM) and $M_{\rm ms}$ ($=$ 8 \SM)
are the upper and lower mass limits of the stars
that explode as Type II supernovae.

\begin{figure}
\plotone{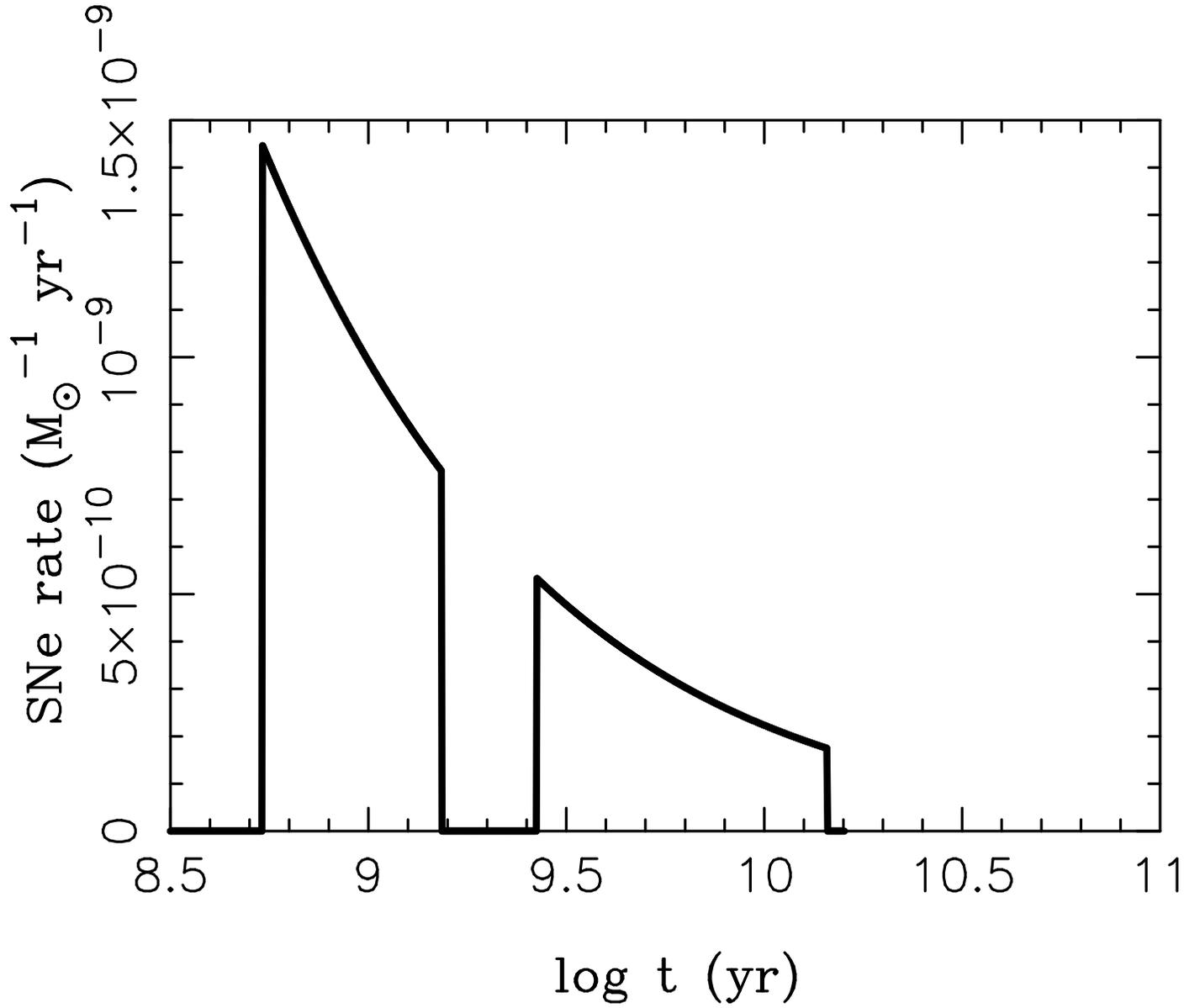}
\caption{
The Type Ia supernova rate ($R_{\rm SNIa}$) as a function of time.
\label{Iarate}
}
\end{figure}

For the Type Ia supernova rate, we adopt the formulation given
in the galactic chemical evolution model by
\citet{Kobayashi_1998, Kobayashi_2000}.
They adopt the single-degenerate scenario \citep{Nomoto_1994}
for the Type Ia supernova progenitor, 
which is that a white dwarf (WD) in a close binary undergoes
a thermonuclear explosion when the companion star evolves
off the main-sequence and transfers a large enough amount of mass over to the WD.
In their model, the Type Ia supernova rate at an epoch $t$ is given as
\begin{equation}
R_{\rm SNIa}(t) = C_{\rm SNIa}
 \frac{\displaystyle{\int^{M(t+\Delta t)}_{M(t)} \phi(m) dm}}{\Delta t},
\label{R_SNIa}
\end{equation}
where $M(t)$ is the mass of the companion star whose main-sequence
lifetime is $t$ and $C_{\rm SNIa}$ is a constant to be calibrated
by the observational constraints.
From the evolution model of the progenitor stars, 
\citet{Hachisu_1996, Hachisu_1999} obtained the mass ranges
of the companion stars of the WDs that become Type Ia supernovae as
\begin{equation}
M_{\rm sec} = [1.8, 2.6], \quad [0.9, 1.5] \quad \mbox{\SM}.
\end{equation}
We note that these mass ranges weakly depend on the metallicity of the star.
In the present paper, we don't adopt the metallicity dependence.
Figure \ref{Iarate} shows $R_{\rm SNIa}$ as a function of time,
where we set $C_{\rm SNIa} =$ 1.0.
In the present paper, we set $C_{\rm SNIa} = 4.0 \times 10^{-4}$
from the result of test calculations.
Note that other authors \citep{Raiteri_1996, Carraro_1998}
adopt different progenitor models by
\citet{Greggio_1983} and \citet{Matteucci_1986}.

For the supernova energy, we assume that
$e_{\rm SNII} = e_{\rm SNIa} = 10^{51}$ erg.
For the stellar wind energy,
$e_{\rm SW}$ is estimated to be 0.2$\times 10^{51}$ erg 
for solar metallicity stars from the observational data of OB associations
\citep{Abbot_1982}.
The chemical abundance of a massive star significantly affects $e_{\rm SW}$ 
\citep{Leitherer_1992}, so that we use metallicity dependent
e$_{\rm SW}$ as $e_{\rm SW} = 0.2 e_{\rm SNII} (Z/Z_{\odot})^{0.8}$, 
where $Z$ is the mass fraction of heavy elements in the STAR.

\subsection{Mass ejection}
In our code, the mass ejection in stellar winds from massive stars
($m \ge M_{\rm ms}$) is combined with the mass ejection
due to Type II supernovae.
The mass ejection in stellar winds from low mass stars
($m < M_{\rm ms}$) is treated separately.
Thus, the mass ejection rate per STAR is written as
\begin{equation}
M_{\rm eject} = m_{\rm SNII} R_{\rm SNII} + m_{\rm SWm} R_{\rm SWm}
	 + m_{\rm SNIa} R_{\rm SNIa}, 
\end{equation}
where $m_{\rm SNII}$ is the average ejecta mass of Type II supernovae,
and $m_{\rm SWm}$ is the average mass ejected in stellar winds
from the low mass stars (corresponding to planetary nebulae),
and $m_{\rm SNIa}$ is the ejecta mass of a Type Ia supernova.
$R_{\rm SWm}$ is the number of stars per unit time expelling their envelopes at a certain epoch.
$m_{\rm SNII}$ and $m_{\rm SWm}$ are defined as
\begin{equation}
m_{\rm SNII} = \frac{\displaystyle{\int^{M_{\rm ma}}_{M_{\rm ms}} m\phi(m) dm}}
 {\displaystyle{\int^{M_{\rm ma}}_{M_{\rm ms}} \phi(m) dm}} - m_{\rm NS}, 
\end{equation}
\begin{equation}
m_{\rm SWm} = \frac{\displaystyle{\int^{M_{\rm ms}}_{M_{\rm ll}} m\phi(m) dm}}
 {\displaystyle{\int^{M_{\rm ll}}_{M_{\rm ms}} \phi(m) dm}} - m_{\rm WD}.
\end{equation}
Here $m_{\rm NS}$ is the mass locked up in a neutron star and
$m_{\rm WD}$ is the mass that is locked up in a WD.
$M_{\rm ll}$ ($=$ 1 \SM) is the mass of the star
whose life time nearly equals to the Hubble time.
We set $m_{\rm NS} =$ 1.4 \SM~ and $m_{\rm WD} =$ 1.4 \SM~
and assume that a Type Ia supernova is produced
by a Chandrasekhar mass WD so that we set $m_{\rm SNIa} =$ 1.4 \SM.
We define $R_{\rm SWm}$ as
\begin{equation}
R_{\rm SWm} = \frac{\displaystyle{\int^{M_{\rm ms}}_{M_{\rm ll}} \phi(m) dm}}
 {\displaystyle{\tau(M_{\rm ll}) - \tau(M_{\rm ms})}}.
\end{equation}
The fraction of heavy elements in $m_{\rm SNII}$ and $m_{\rm SNIa}$
is computed using the nucleosynthesis yield of Type II and Ia supernovae
\citep{Tsujimoto_1996, Nomoto_1997}.
We compute the chemical evolution of total metal (Z), iron (Fe) and
oxygen (O) in our code.
The metallicity yield we used is tabulated in Table \ref{yield}.

\subsection{Summary on Feedback}
\begin{figure}
\plotone{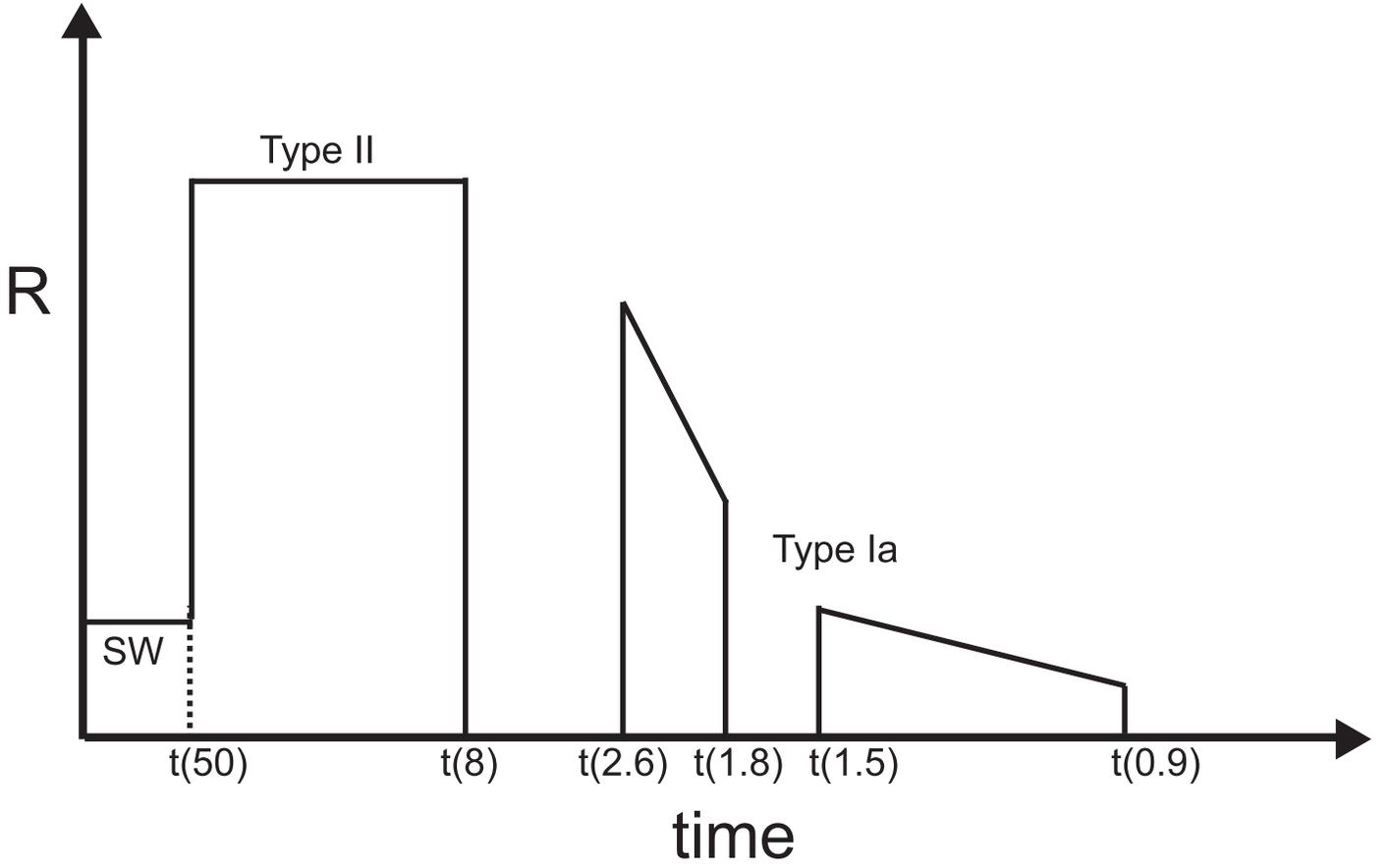}
\caption{
The schematic view of the feedback processes in our SPH code.
``t(m)'' is the stellar lifetime as a function of main-sequence mass (m/\SM).
\label{feedback}
}
\end{figure}

The feedback phase is divided into three phases;
a stellar wind phase, a Type II supernova phase,
and a Type Ia supernova phase.
\begin{itemize}
\item
The stellar wind phase continues for $\tau(M_{\rm ma})$, 
during which only the energy ejection from STARs is included;
the ejected mass is included in the Type II supernova phase for simplicity.

\item
The Type II supernova phase begins at $t = \tau(M_{\rm ma})$ and
ends at $t = \tau(M_{\rm ms})$.
During this phase, the mass ejection rate is the sum of
the contributions by stellar winds of massive stars and Type II supernovae.

\item
The Type Ia supernova phase begins at $t = \tau(M_{\rm ms})$.
During this phase, both the energy ejection and mass ejection 
rates are the sum of the contributions by Type Ia supernovae
and stellar winds of low mass stars.
\end{itemize}
We present the schematic view of the feedback processes
in Figure \ref{feedback}.

With the present algorithm and the adopted parameters,
a STAR of 10$^8$ \SM~ produces $\sim 5.5 \times 10^5$ Type II
supernova explosions and $\sim 1.4 \times 10^5$ Type Ia supernova explosions
during 13 Gyr of the evolution.
Also $\sim 2.2 \times 10^7$ \SM~ gas is ejected during the evolution.
The ejected gas contains $\sim 1.6 \times 10^6$ \SM~ of heavy elements, 
which include $\sim 1.5 \times 10^5$ \SM~ of iron and
$\sim 1.0 \times 10^6$ \SM~ of oxygen.

As a final note, the thermal energy, gas, and heavy elements from stellar winds and
supernovae are smoothed over neighbor particles of
the STAR within a neighbor radius of $R_f$ (a feedback radius).
$R_f$ is a parameter set to 0.5 kpc,
which is equal to the adopted gravitational softening
length of STAR particles.
Results of a test calculation with $R_f =$ 0.8 kpc are
very similar to those of the present paper.

\section{Initial model}
Following the previous work \citep[e.g.,][]{Katz_1992}, 
we investigate the evolution of the spherical 3 $\sigma$ top hat over-dense
region of a mixture of dark matter and gas with our SPH code.
We neglect the matter outside the spherical regions,
thus neglecting the late in-fall of the external matter and the
external gravitational field.
Throughout the paper, we set the Hubble constant 
$H_0=$ 50 km s$^{-1}$ Mpc$^{-1}$, 
$\Omega =$ 1, and $\Lambda =$ 0.
We set the initial redshift ($z$) of the model to be $\sim$ 25
and the initial radius of the sphere to be 55 kpc.
Thus, the comoving radius of the spherical region is $\sim$ 1.4 Mpc
and the contained mass is $\sim 10^{12}$ \SM.
We set the region in rigid rotation to provide a sufficient angular momentum
with the spin parameter $\lambda \sim 0.1$.
We also add the corresponding Hubble velocity to the velocity field of the sphere
since we integrate the model with the physical coordinate.

To generate a 3 $\sigma$ over-dense region, 
we use the path integral method \citep{Bertschinger_1987}.
We note that there are further two methods to set up the over-dense region:
(1) To seek the desired over-dense region by sampling different random
field realizations, which is used by the cosmological hydrodynamical
simulations of formation of X-ray clusters \citep{Anninos_Norman_1996}.
(2) To select the desired halo from the results of the large scale
cosmological N-body simulations and use it as an initial condition for
a cosmological galaxy formation model \citep{Navarro_White_1994}.
The path integral method is the fastest method among other methods.
And it is the easiest way to generate a several dozen desired over-dense regions,
i.e. the over-dense region of 10$^{12}$ \SM~ in the present study.
The weakness of this method is that we have to use the isolated boundary condition.
Namely, we neglect the external tidal field and
artificially add the angular momentum to the initial velocity field.
On the other hand, with the method (2), it is natural to use
the surrounding particles as the origin of external tidal-field.
However, with the method (2), it is difficult to find several over-dense regions
with similar properties since each region is very different like the real universe.
We think our way to generate initial conditions is 
an efficient way for our purpose to construct the model for our Galaxy.

Specifically, initial models are constructed as follows:
(1) We generate a spherical top hat 3 $\sigma$ over-dense region
by the path integral method \citep{Bertschinger_1987}.
We use the COSMICS package \citep{Bertschinger_1995}
and generate 50 realizations.
(2) We follow the non-linear evolution of the dark matter particles
in each spherical region with the same code.
Note that we only use the N-body part of our GRAPE-SPH code
in these simulations.
(3) At a certain epoch ($z \sim$ 2.5),
we examine the properties of dark matter halos and
select an appropriate realization for the hydrodynamical model.
We select the model with a single halo since if two dominant halos exist,
the gas disk will be destroyed by a major merger event \citep{Barnes_1992}.
Among 50 realizations, 39 realizations are single halo models and 
others are multiple halo models.
From 39 realizations, we select candidates for the hydrodynamical model
by comparing the model density profile with the density profile of the Galactic halo.
(4) Once we select several models, 
we restart the hydrodynamical N-body simulation from $z \sim$ 25
with our GRAPE-SPH code.
After we have done full hydrodynamical calculations for three single-halo models
(and two multiple-halo models),
we select the best-fit model, which is presented in this paper, 
by comparing the disk scale length with the Galactic scale length.
All results shown in the following sections are obtained with this best-fit model.

We have computed a similar kind of models
with several different configurations
and found that the performance mainly depends on the speed of the GRAPE system:
(a) For our old configuration using four GRAPE-3A boards,
a similar calculation to the present work
takes more than 200 hours up to $z \sim$ 0.9 ($t =$ 5 Gyr). 
(b) For our recent configuration using one GRAPE-5 board, 
a similar calculation takes only $\sim$ 150 hours up to $z =$ 0 ($t =$ 13 Gyr).
Using this recent configuration, 
a full hydrodynamical N-body computation for several dozen realizations
are under way and results will be presented elsewhere.

\section{Formation of the Galaxy}
The initial number of gas particles and dark matter particles 
are $\sim 27,000$ and $\sim 27,000$, respectively.
Thus, with the total mass of $10^{12}$ \SM, 
the initial masses of the gas particles and dark matter particles
are $4 \times 10^6$ \SM~ and $3.6 \times 10^7$ \SM, respectively.
We set the gravitational softening length for
the gas and dark matter particles to be 0.5 kpc and 1.0 kpc, respectively. 
The overall evolution is very similar to the previously reported work
\citep[][]{Steinmetz_Muller_1995, Berczik_1999}.
During the expansion due to the Hubble flow, small scale structures grow and
dense regions collapse.
In such high density regions (clumps), star formation occurs as a result of 
the efficient radiative cooling.
These small clumps gradually merge to produce larger clumps.
At $t \sim$ 0.3 - 0.5 Gyr, these clumps merge and
the resulted clump becomes
the primary halo that eventually becomes a spiral galaxy.
Around this time ($z \sim$ 7), the star formation rates (SFR)
reaches a maximum value.
At $t \sim$ 1 Gyr ($z \sim$ 4), the dark matter is almost virialized and
the evolution of dark matter particles becomes a quasi-static state while
the gas particles are being settled into a thin disk.
Finally, the prominent gas disk forms in the center of the primary halo
at $t \sim$ 3 Gyr ($z \sim$ 1.6).
Afterward, the most star formation occurs in this gas disk and
the overall evolution becomes a quasi-static state.

After 5 Gyr of evolution (up to $z \sim$ 0.9),
over 80,000 STARs have formed and
the number of gas particles have decreased to $\sim 13,000$.
At this stage, we categorize STARs according to
the chemical properties and the formation epoch ($t_{\rm form}$)
of STARs following the observational characteristics of the three components
of the Galaxy as defined in \citet{Mihalas_1981}.

\begin{figure}[h]
\plotone{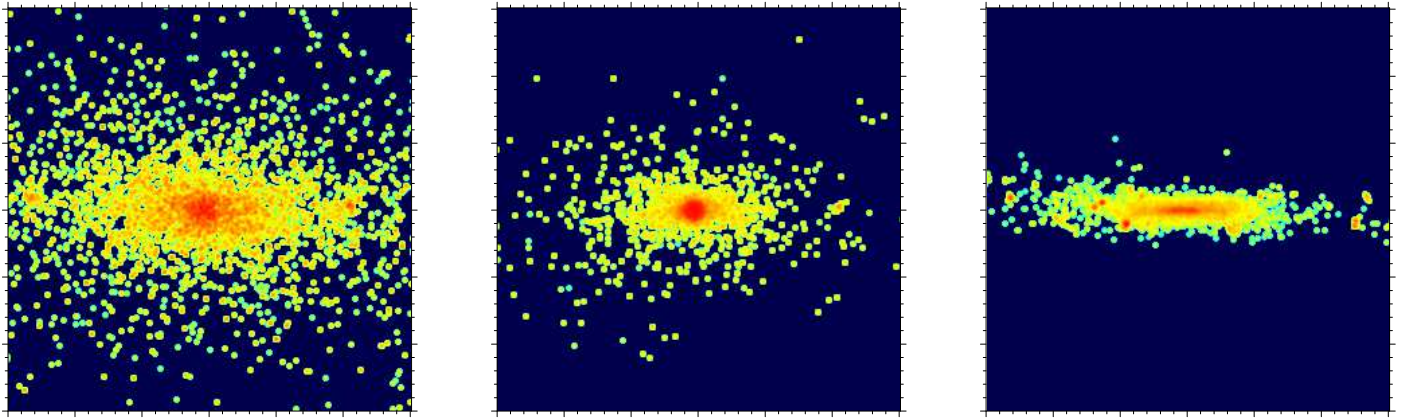}
\caption{
The projected STAR positions at $t =$ 5 Gyr ($z \sim$ 1)
for the three components.
From the left panel, the metal-poor STARs, the metal-rich and old STARs,
and the young STARs are shown (see text).
The size of the panels is 20kpc $\times$ 20kpc.
\label{3comp}
}
\end{figure}

Specifically, we categorize STARs by the following conditions and
show the projected stellar densities in Figure \ref{3comp}:
\begin{itemize}
\item left panel : Metal-poor (log(Z/Z$_{\odot}$) $< -3$) STARs,

\item center panel : Old ($t_{\rm form}$ $< 1$ Gyr) and
metal-rich (log(Z/Z$_{\odot}$) $> 0$) STARs,

\item right panel : Young ($t_{\rm form}$ $> 4$ Gyr) STARs.
\end{itemize}
Here, $Z$ is the mass fraction of heavy elements in STARs.
Metal-poor STARs show an extended distribution since these STARs 
formed at an early epoch and at high latitude (the left panel).
The old and metal-rich STARs are concentrated in the galactic center
(the center panel).
The qualitative properties of the system are
not so changed after $z \sim$ 1 as reported by \citet{Steinmetz_Muller_1995}.
Since we neglect the late in-fall of surrounding material,
the most recent star formation mainly occurs in the disk.
As a result, the young STARs are located near the galactic plane
(the right panel).
These results show that the calculated stellar system 
looks very similar to the Galaxy.
To be more qualitatively, we plot the face-on surface density profile
of STARs in Figure \ref{density}. 
The solid crosses show our result. 
We fit the model with the following function;
$\rho(r) = \rho_{\rm disk} {\rm exp}(-r/R_{\rm disk}) + \rho_{\rm bulge} {\rm exp}(-7.67((r/R_{\rm bulge})^{0.25}-1))$, 
where the scale radiuses are $R_{\rm disk} =$ 4.21 kpc and $R_{\rm bulge} =$ 0.876 kpc.
In the next section, we analyze the properties of STARs
near the galactic center.

At $z =$ 0, the total number of particles has increased
to $\sim$ 150,000 where the numbers of gas, dark matter, and STAR particles
are $\sim$ 11,000, 27,000, and 112,000, respectively.
When we compare the projected density profile at $z \sim$ 0.9 (Figure \ref{3comp})
and at $z =$ 0, the 3 components structure looks very similar
except the thickness of the disk STARs.
Most star formation occurs quasi-statically near the disk plane and
the SFR is $\sim$ 1 - 2 \SM~ yr$^{-1}$.

\section{Formation history of the bulge component}
The one-zone chemical evolution model \citep[e.g.,][]{Matteucci_1999}
predicted the approximate chemical properties of the Galactic bulge.
To construct a self-consistent model of the bulge formation,
however, a chemical and dynamical model is required.
In this section, we use our high resolution model for the 
formation and evolution of the Galaxy
to study the formation of the Galactic bulge.

\subsection{Dynamical history and kinematics of the bulge stars}
We select the STARs located near the galactic center ($R <$ 2 kpc)
and define these STARs as the bulge stars.
We define the center of the gravity of all STARs as the galactic center.
To find the galactic center, we need several times of a iteration
to compute the center of the gravity.
In the $i$-th iteration, we compute the center of the gravity of STARs
that is located within $R_i$ kpc from the origin of the coordinate
and convert the origin with that point.
We set $R_1 =$ 20 kpc and $R_2, R_3, ... =$ 5 kpc.
Typically, we only need three or four iterations.
For example, at $z \sim$ 0.9, the mass and number of the bulge stars are
$1.8 \times 10^{10}$ \SM~ and 16,800, respectively.
Ever after this time, we can select the bulge stars with the same definition.
In this paper, however, we compare our model results at this time
with some observational data.
The reason is that in SPH models including star formation recipes, 
the resolution of the model becomes worse and worse during the evolution
since the number of gas particles decreases.
With our chemical enrichment model,
we need to have a large enough number of gas particles
around a STAR particle to properly model the chemical evolution
of the interstellar matter;
further, the gas particles show a more extended distribution
than the star particles and this distribution makes the effective resolution
of the chemical enrichment description worse.
At $z \sim$ 0.9, the number of gas particles becomes $\sim$ 13,300, 
which is a half of the initial number of gas particles
and sufficiently large.

\begin{figure}
\plotone{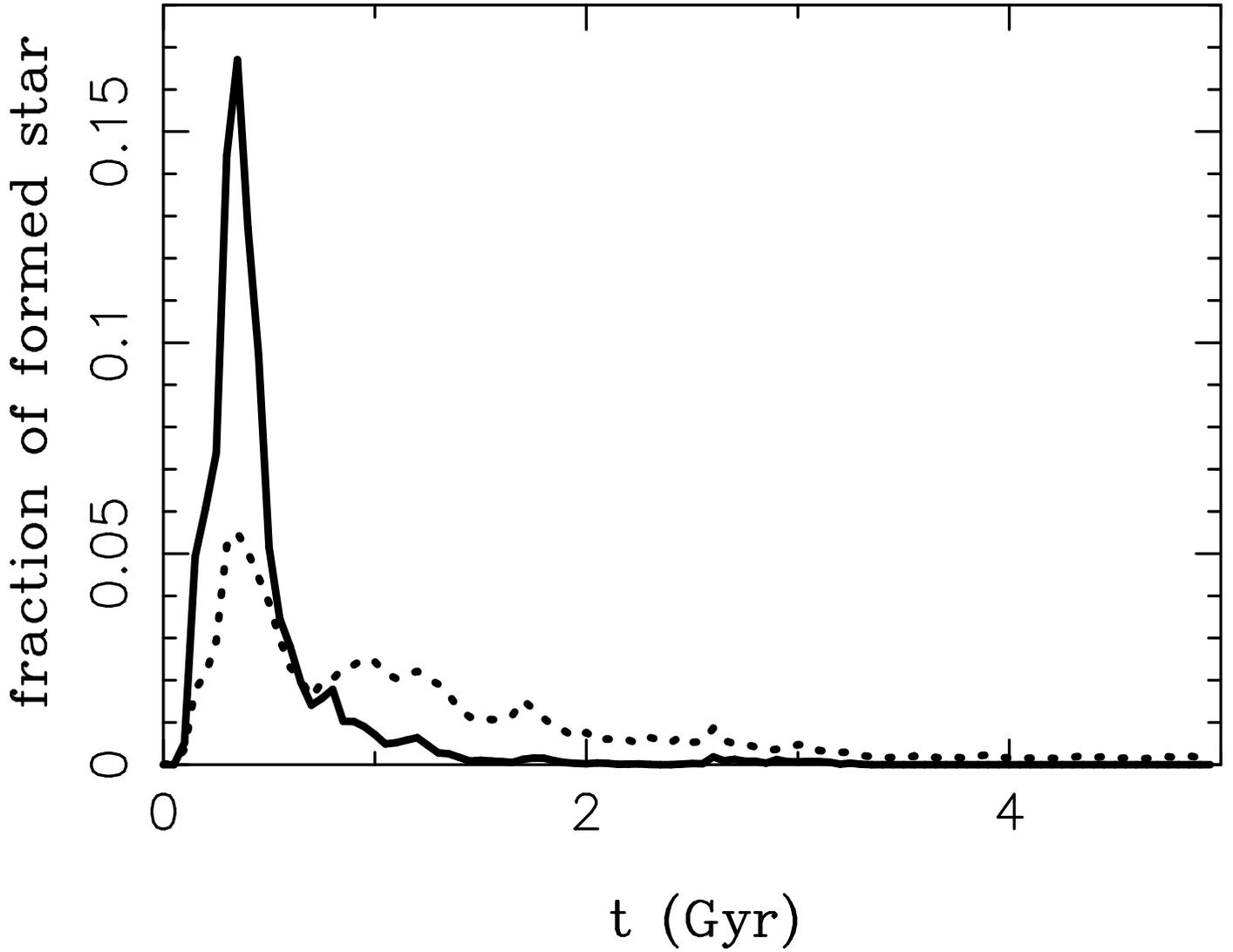}
\caption{
The star formation history for the bulge stars (solid line)
and all stars (dotted line).
The SFR divided by the total stellar mass of each population
is shown as a function of time.
\label{star_sff}
}
\end{figure}

Figure \ref{star_sff} shows the SFR
as a function of time for the bulge stars (solid line)
and all stars (dotted line) up to $z \sim$ 0.9.
In this figure, we normalize the SFR
by the total mass of each component at $z \sim$ 0.9.
More than 60 \% of the bulge stars up to this time
formed during the first 0.5 Gyr of the evolution.
With our model, we can obtain the detailed formation history of the bulge stars.
In Figure \ref{bulge_p}, we present snap shots for
projected STAR positions in the region where star formation
is most violent for the first 0.5 Gyr.
We note that the merging of two sub-galactic clumps occurs during
$t =$ 0.4 - 0.5 Gyr.
This merging of the clumps causes the star burst of 
the short time scale shown in Figure \ref{star_sff}.
Such a phenomenon is different from the simple star burst
assumed in the one-zone chemical evolution model
\citep{Matteucci_Brocato_1990}.

With our chemical and dynamical model, 
we can also obtain the kinematics of the bulge stars.
In Figure \ref{bulge_vel}, we plot the average velocity (upper line)
and the velocity dispersion (lower line) of these bulge stars.
The bulge stars are substantially rotationally supported
with the rotational speeds of $\sim$ 160 - 230 km s$^{-1}$.
These speeds are smaller than the rotational speeds of the disk stars
($\sim$ 270 km s$^{-1}$ in the present model), which means that
the angular momentum of the bulge is lost during the merging process.
Figure \ref{bulge_dis} shows the x-direction velocity dispersion
for the bulge stars as a function of [Fe/H].
Here, our numerical results are shown by the solid triangles and solid squares
and compared with observational data \citep{Minniti_1996} shown by the doted circles
(note that the observational data is the line of sight velocity dispersion).
The solid squares are the STARs with $t_{\rm form} < 1$ Gyr (old STARs) and
the solid triangles are the STARs with $t_{\rm form} < 5$ Gyr (young STARs).
The old and young STARs show a different trend,
and the result for old STARs is more consistent with the observational data.
Our model (especially old STARs) qualitatively reproduces
the kinematic properties of the Galactic bulge.

\subsection{Chemical properties of bulge stars}
Important information to constrain the formation of
the Galactic bulge concerns the chemical properties of the bulge stars.
In this subsection, we analyze the chemical properties of the bulge stars.

The observed metallicity distribution function of K-giant stars 
\citep{McWilliam_Rich_1994} is characterized by 
a wide ($-1.2 <$ [Fe/H] $< 0.9$) distribution
as shown by the dashed line in Figure \ref{bulge_mdist}.
The distribution function of the bulge stars in our model
(the solid line in Figure \ref{bulge_mdist}) is
as wide as $-2.0 <$ [Fe/H] $< 1.0$, being in good agreement
with the observational data. 
As noted in the previous subsection, about 60 \% of the bulge stars
form in the strong star-burst caused by merging
of sub-clumps (hereafter these stars are called a merger component).
This merger-driven star-burst is partly responsible for 
this wide distribution.
On the other hand, other 40 \% of the bulge stars 
form in the late stage of evolution (a non-merger component)
and also contribute to widen the distribution function.
This is because the non-merger component of the bulge stars
has much larger [Fe/H] than the merger component as explained below.

The bulge stars can be divided into two chemically different types
with and without being contaminated by Type Ia supernova ejecta.
With the adopted progenitor model \citep{Kobayashi_1998}, 
the epoch of the first appearance of Type Ia supernovae
$t_{\rm SNIa} =$ 0.5 Gyr after the formation of each STAR.
Thus, the STARs that formed after $t =$ 0.5 Gyr
are partially contaminated by Type Ia supernovae.
Each Type Ia supernova ejects $\sim 10$ times more iron than
a Type II supernova \citep{Nomoto_1997}.
As a result, iron contained in the non-merger component of the bulge stars
partially originates from Type Ia supernovae and their [Fe/H] tends to
be higher than that of the merger component.
In Figure \ref{bulge_time}, the distribution function of the STARs
that formed before and after $t =$ 0.5 Gyr are plotted by the dashed and dotted lines,
respectively, along with the total distribution function of the bulge stars
(solid line).
Both dashed and dotted lines show rather wide distribution
and clearly the peak value of each distribution is different.
The very old STARs, which form before $t =$ 0.5 Gyr, 
contribute mainly to the metal poor range of the distribution.

The adopted yields are [O/Fe] $=$ 0.4 from Type II supernovae and 
[O/Fe] $= -1.5$ from Type Ia supernovae (Table \ref{yield}).
Thus the contamination by Type Ia supernovae ejecta decreases
[O/Fe] in STARs starting from [O/Fe] $= 0.4$.
In Figure \ref{bulge_o_fe}, we present the metallicity distribution
function of the bulge stars for [O/Fe].
The very peaky distribution shows that 
a half of the bulge stars have [O/Fe] $\sim 0.4$, 
i.e. a half of the bulge stars are not contaminated by Type Ia supernovae.
Also, almost 80 \% of the bulge stars have [O/Fe] $> 0$.
Figure \ref{bulge_o_fe_obs} shows [O/Fe] as a function of [Fe/H]
for the bulge stars.
[Fe/H] increases up to $\sim 0.3$ without Type Ia supernova contamination.
Later, due to the increasing Fe enrichment by Type Ia supernovae, 
[O/Fe] decreases down to $\sim -0.4$
as [Fe/H] increases to $\sim 0.7$.
We note for a given  [Fe/H] ($> -0.5$), [O/Fe] is distributed over
a wide range of $> -0.4$.
Thus, even for [Fe/H] $> 0$, large fraction of the bulge stars
have [O/Fe] $> 0$, while [O/Fe] is mostly sub-solar for [Fe/H] $> 0.3$.
These features of our model are consistent with the observed
results as plotted in Figure \ref{bulge_o_fe_obs}
by the dotted circles \citep{Rich_2000}.
The observed [Si/Fe] and [Ca/Fe] show also relatively large
dispersion from $-0.4$ to $+0.2$ for [Fe/H] $> 0$, 
although the observed results are still preliminary
\citep{Rich_2000}.

In contrast, \citet{Matteucci_Brocato_1990} argued
that the metallicity of the bulge stars mainly originated
from Type II supernovae.
In their model, they assumed $t_{\rm SNIa} \sim$ 1 Gyr and
the star formation time scale is shorter than 1 Gyr
so that the majority of Type Ia supernovae have exploded
long after the star formation.
Therefore the one-zone chemical evolution model predicts
that the Galactic bulge is a chemically single population.
In contrast to this picture, our chemical and dynamical model predicts
that the Galactic bulge is neither a chemically single population
and a dynamically single population.

\subsection{Discussion}
There are several problems in our numerical model.
First, the present model, which is compared with some observational data, 
is just the current best-fit model from only 5 realizations.
Thus, the quantitative comparisons between
our model and observation are still far from complete.
One reason is that it is too difficult to find a consistent
initial model for very good match to the observation
before doing real calculations.
Due to complicated merging events occurring in the early phase, 
it is impossible to predict which initial halo
will become a Milky-Way-like spiral galaxy. 

\begin{figure}
\plotone{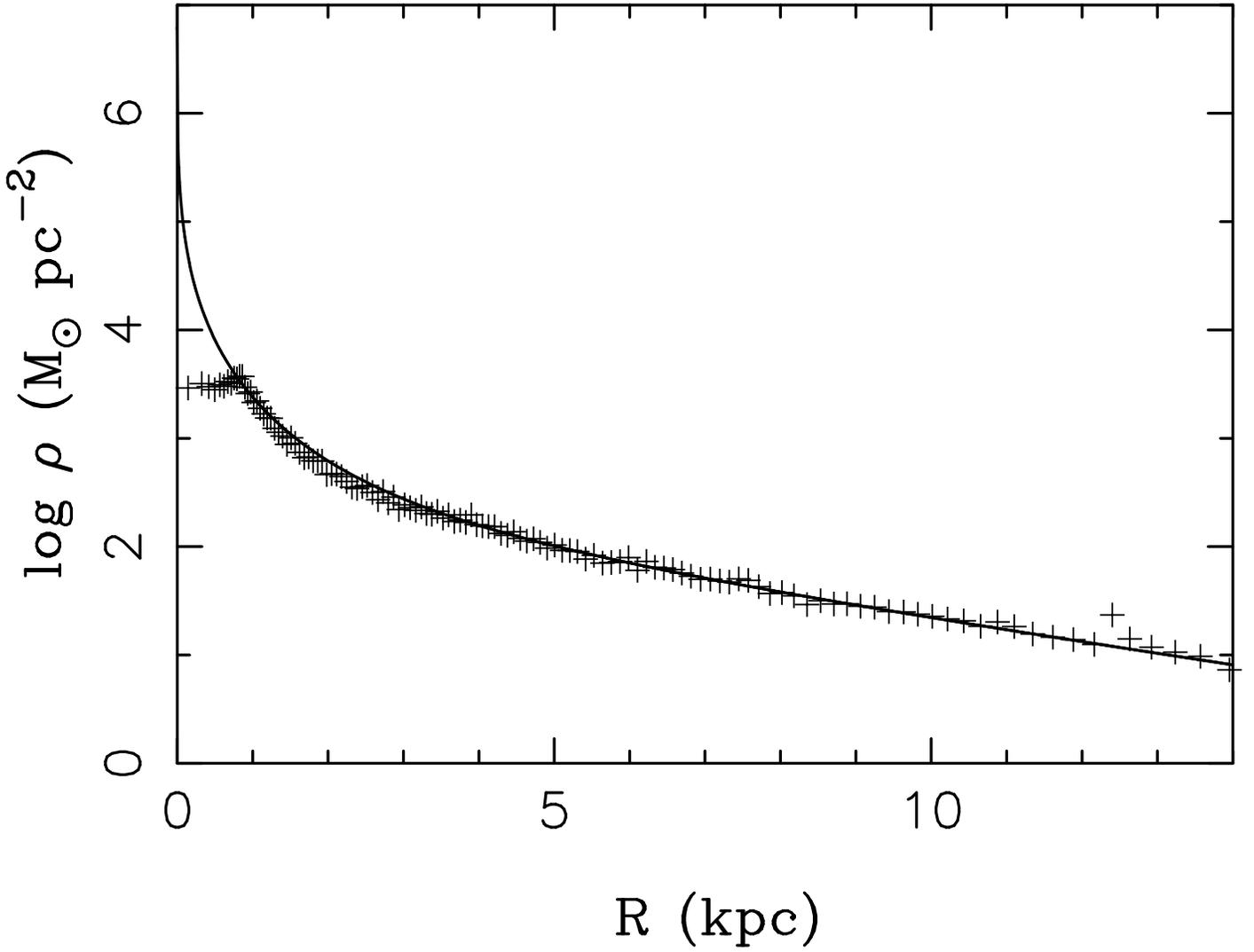}
\caption{
The face-on surface density of STARs at $t =$ 5 Gyr.
The crosses represent our model and 
the solid line is the fit to our model (see text).
\label{density}
}
\end{figure}

\begin{figure}[h]
\plotone{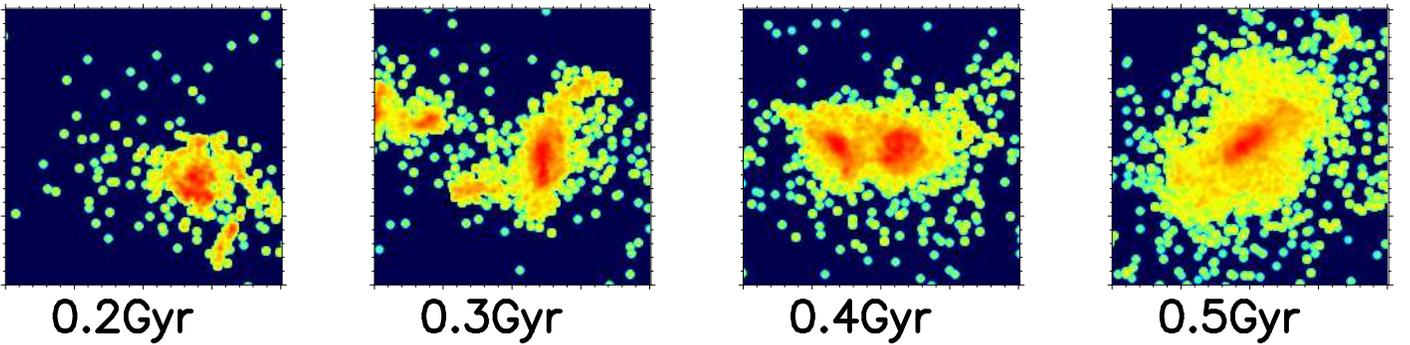}
\caption{
The projected STAR position for the first 0.5 Gyr of the evolution.
During $t =$ 0.4 - 0.5 Gyr, two clumps merge together.
The size of the panels is 10kpc $\times$ 10kpc.
\label{bulge_p}
}
\end{figure}

\begin{figure}
\plotone{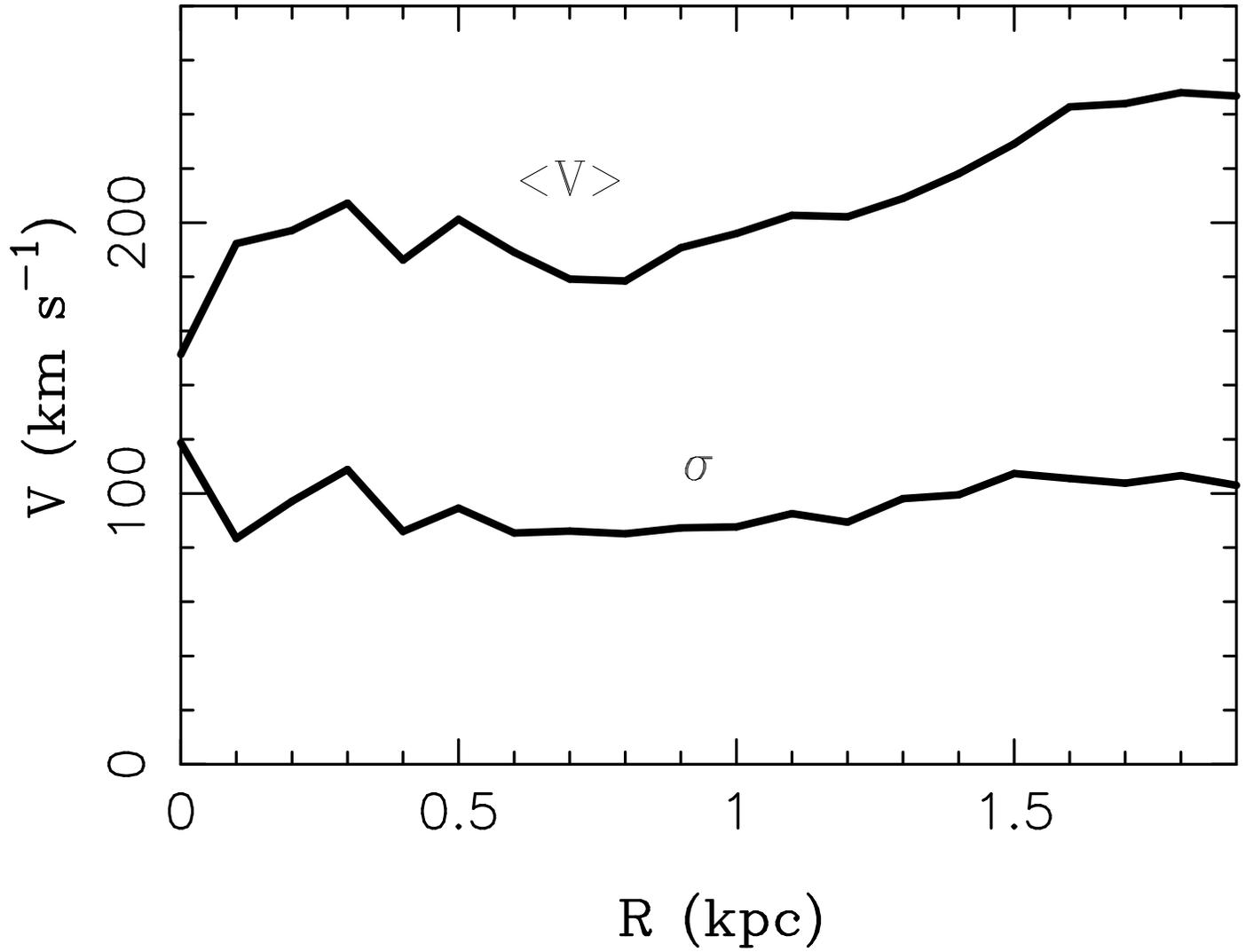}
\caption{
The mean velocity and the velocity dispersion for the bulge stars
as a function of the distance from the galactic center.
The upper and lower lines are the average velocity and the velocity dispersion, 
respectively.
\label{bulge_vel}
}
\end{figure}

\begin{figure}
\plotone{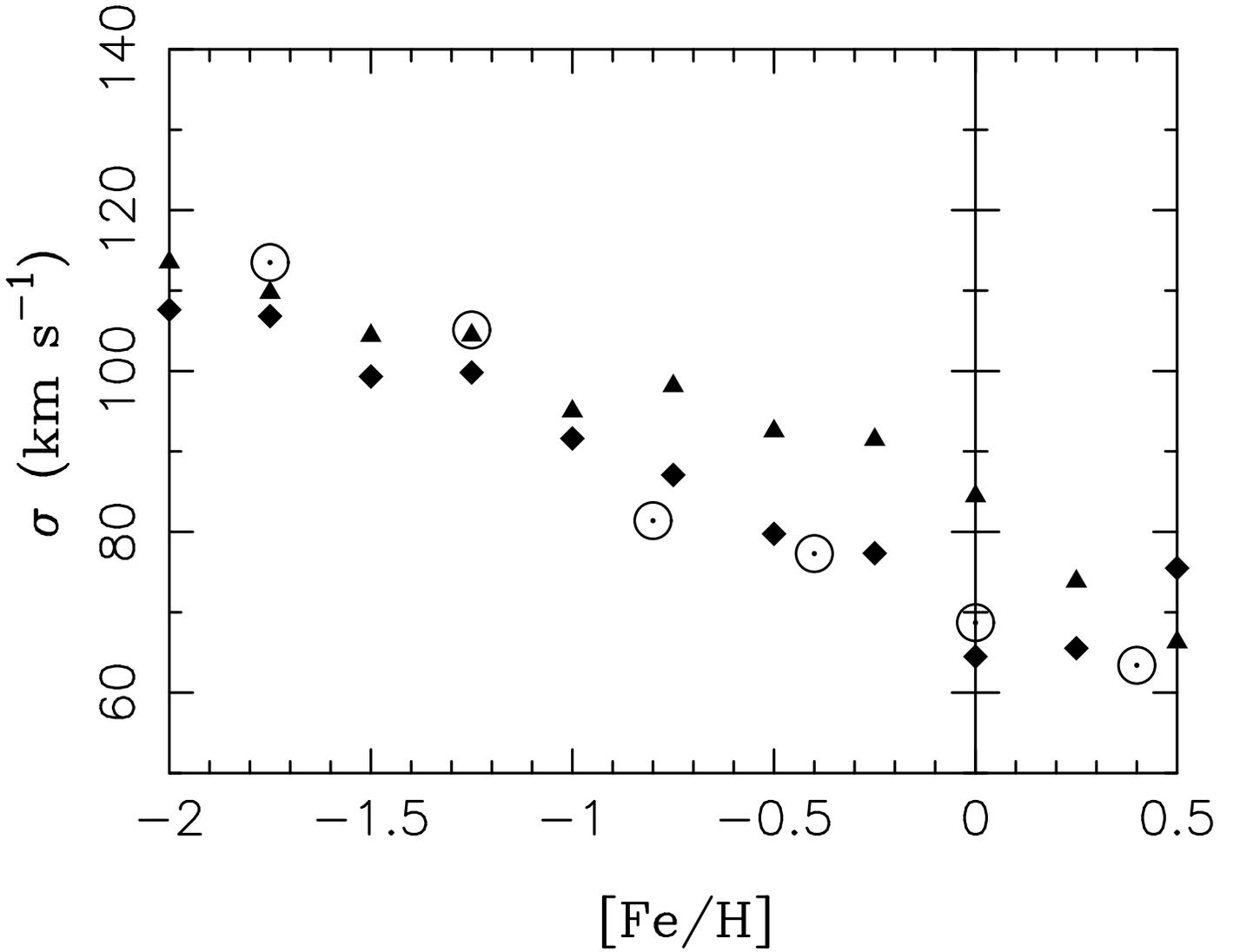}
\caption{
The x-direction velocity dispersion for the bulge stars
as a function [Fe/H].
The dotted circles are the observational data in \citet{Minniti_1996}
(the error bound in the observation is $\sim$ 8 - 14 km s$^{-1}$).
Our model STARs are shown by the filled squares ($t_{\rm form} < 1$ Gyr)
and the filled triangles ($t_{\rm form} < 5$ Gyr).
\label{bulge_dis}
}
\end{figure}

\begin{figure}[h]
\plotone{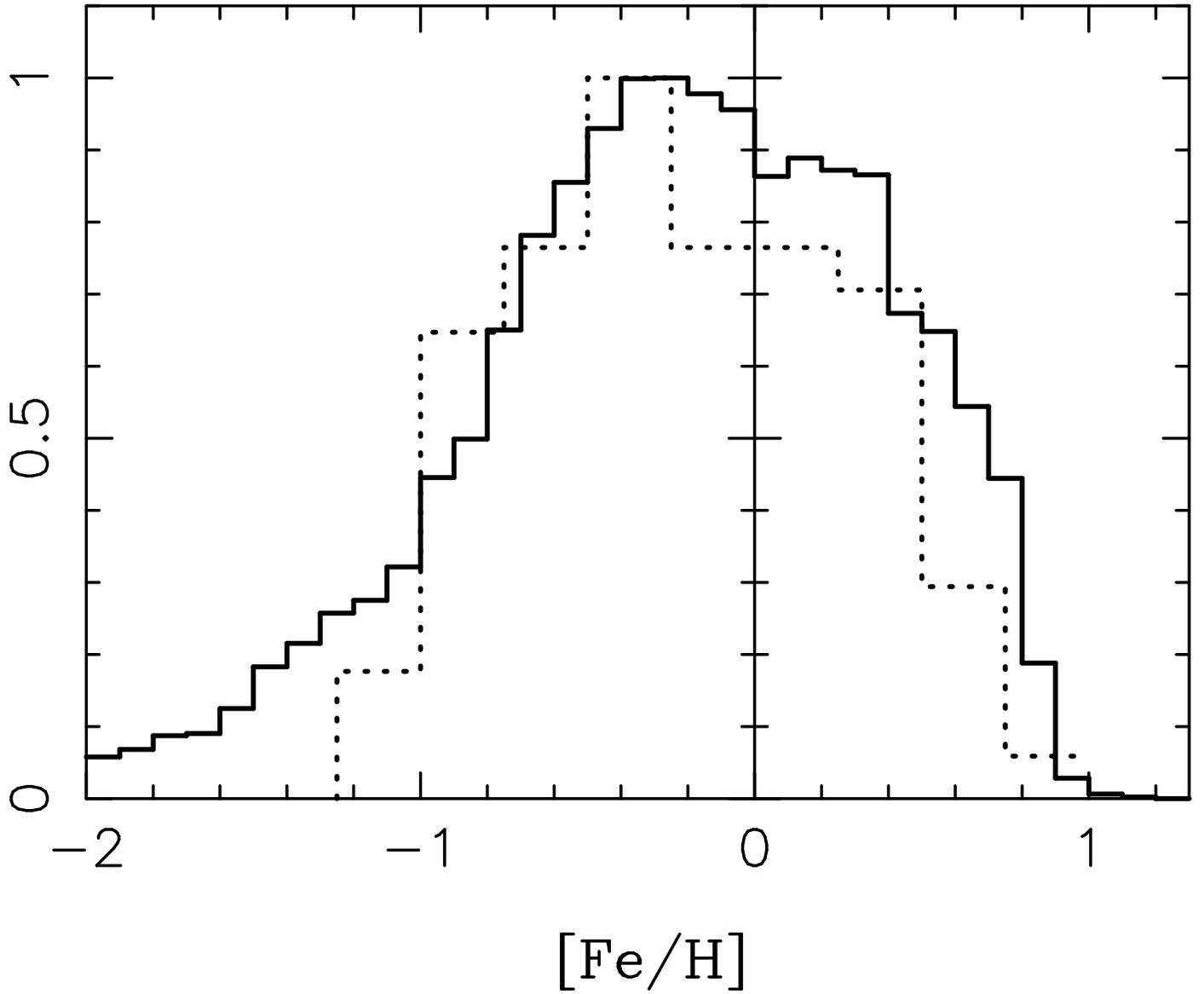}
\caption{
The calculated metallicity ([Fe/H]) distribution function of the bulge stars (solid line).
The dashed line shows the observed metallicity distribution function of K-giant stars 
\citep{McWilliam_Rich_1994}.
The distributions are normalized by its maximum value for comparison.
\label{bulge_mdist}
}
\end{figure}

\begin{figure}
\plotone{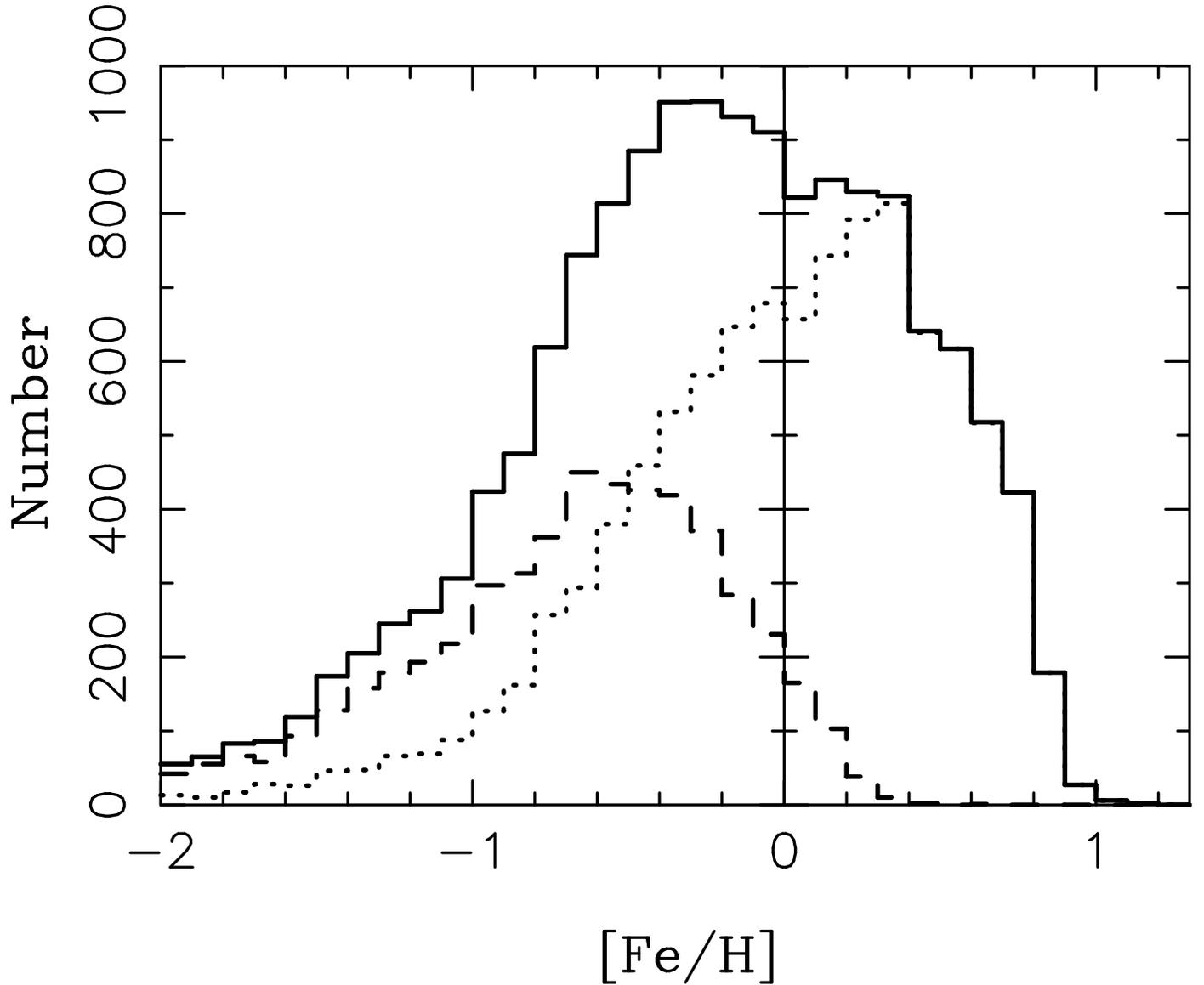}
\caption{
The calculated metallicity ([Fe/H]) distribution function of the bulge stars (solid line).
The distributions are not normalized.
The dashed and dotted lines correspond to the very old STARs with $t_{\rm form} < 0.5$ Gyr
and young STARs with $t_{\rm form} \geq 0.5$ Gyr, respectively.
\label{bulge_time}
}
\end{figure}

\begin{figure}
\plotone{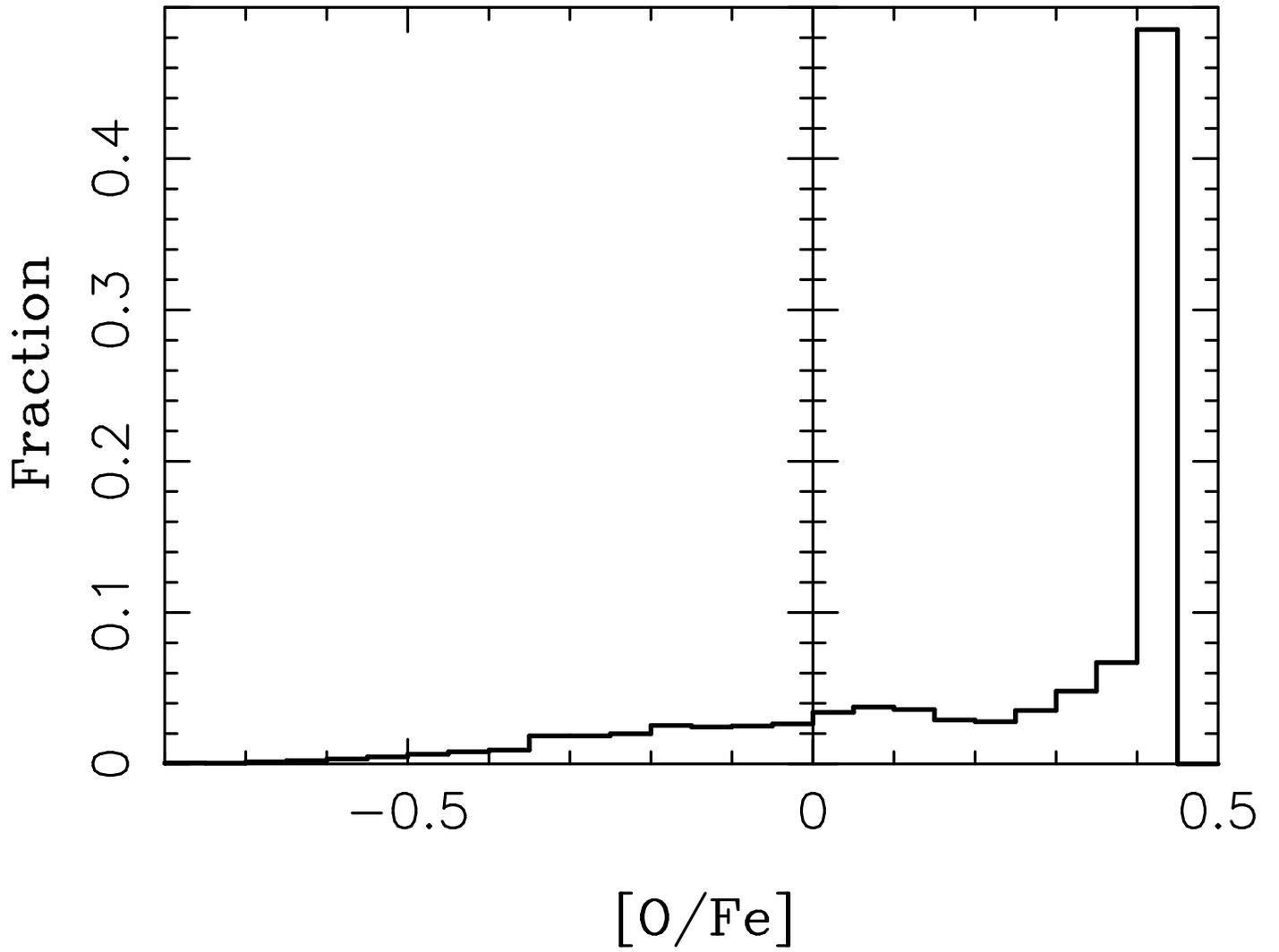}
\caption{
The calculated metallicity ([O/Fe]) distribution function of the bulge stars.
\label{bulge_o_fe}
}
\end{figure}

\begin{figure}
\plotone{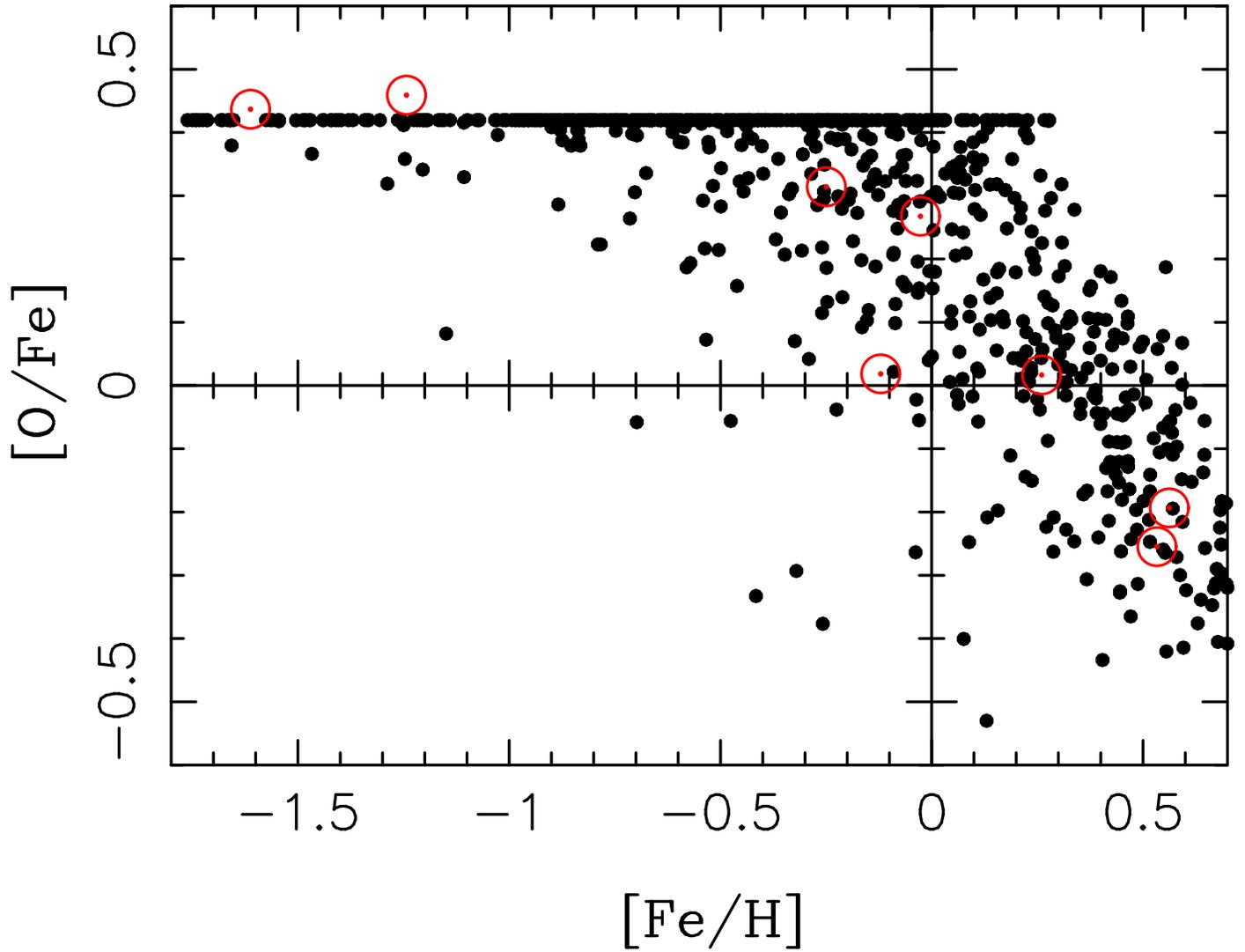}
\caption{
[O/Fe] vs [Fe/H] for $\sim$ 800 STARs, which are randomly selected
from the bulge stars in our model (filled circles).
The dotted circles show preliminary observational data \citep{Rich_2000}.
\label{bulge_o_fe_obs}
}
\end{figure}

Another point is the resolution of the numerical model.
As seen in Figure \ref{density}, the inner region 
of the present model is not well resolved.
Thus, our definition radius for the bulge, $R <$ 2 kpc,
is not a clear definition when we compare our results with observational data.
This is because the Baade's window, where the chemical properties
of the bulge stars are observed, is located 0.5 kpc from the galactic center.
The choice of our definition is determined from the limitation
of the numerical resolution of the present model.
We use the rather large radius of 2 kpc to define the bulge stars
since the gravitational softening length of the present model is 0.5 kpc
(in fact, the inspection of Figure \ref{density}
shows that the effective resolution in our model is $\sim$ 1 kpc, which equals to 
the softening length of the dark matter particles).
To summaries, the number of particles in the present model is
still too small to fully resolve the chemical structure of the bulge.
To make more sensible comparison between numerical models and the observational data,
we need much higher resolution models
and more observational data obtained with large telescopes.
Although there is such limitation, our model 
is in a good agreement with chemical and kinematic
properties of the Galactic bulge stars.

\section{Conclusion}
In this paper, we report on the result of our high resolution
three-dimensional models of the Galactic bulge formation.
By modeling the formation and evolution of the Galaxy, 
we understand the formation history of the Galactic bulge as follows.

The early evolution of the best-fit model reveals that
most bulge stars form during the merger
that occurred in the region of the deepest potential.
Because of the strong star burst induced by the merger, the metallicity
distribution of the bulge component becomes as wide as observed.
Although such a strong star burst was a similar effect as
the burst assumed in the previous chemical evolution model,
our model show that there is another reason to widen the
metallicity distribution function of the bulge.
By analyzing our results, we find that the bulge stars in our model
are composed of two different components.
One component (merger component) was produced by the merger occurred
in the center of the halo as noted above.
The other non-merger component was formed in the inner region of 
the disk after this merger.
Due to the different formation epoch,
the main source of iron for each component is different.
Iron in the former comes mainly from Type II supernovae, 
and the other from Type Ia supernovae.
Since each Type Ia supernova ejects $\sim 10$ times more iron than
a Type II supernova, [Fe/H] of the non-merger component
tends to be higher than that of the merger component, 
which widened the metallicity distribution function.
Also, our model predicts  a relatively large dispersion
in [O/Fe] for [Fe/H] $> 0$.

Although our model is not in a complete agreement
with our Galaxy, some chemical and kinematic properties of the Galactic bulge, 
such as the velocity dispersion vs. metallicity, are well reproduced.
We thus conclude that an old fraction  of the Galactic bulge is very likely
to have been formed by the sub-clump merger in the proto-galactic environment, 
while a younger fraction of the Galactic bulge has formed gradually in the inner disk.
These two groups should show different chemical properties.

We would like to thank the anonymous referee
for the valuable comments and suggestions to greatly improve the manuscript.
Also, we would like to thank J. Makino for kindly providing
access to the GRAPE-4 and GRAPE-5 systems, 
on which a part of the computation has been done.
Some part of the computation has been performed with
the GRAPE cluster at the Astronomical Data Analysis Center of
the National Astronomical Observatory, Japan.
N.N. specially thanks the Astronomisches Rechen-Institut, Heidelberg
for the hospitality during the stay as a guest researcher
where a part of the paper has been written.
This work has been supported by JSPS Research Fellowships for 
Young Scientists (7664) and (05127), 
in part by an SFB 439 grant at the University of Heidelberg and
by the grant-in-Aid for Scientific Research
(07CE2002, 12640233, 14047206, 14540223)
of Ministry of Education, Science, Culture, Sports and Technology in Japan.

\clearpage
\begin{deluxetable}{cccccc}
\tablecaption{The heavy element yields used in our code
\label{yield}
}
\tablehead{
\colhead{} &
\colhead{M$^{a}$} &
\colhead{Z} &
\colhead{Fe} &
\colhead{O}
}
\startdata
Type II SNe & 14.0 \SM & 2.54 \SM & 0.091 \SM & 1.80 \SM \\
Type Ia SNe & 1.4  \SM & 1.4  \SM & 0.74  \SM & 0.14 \SM \\
\enddata
\tablecomments{
These data are taken from \citet{Tsujimoto_1996} and \citet{Nomoto_1997}.\\
$^{a}$The total ejected mass per one supernova explosion. \\
}
\end{deluxetable}

\begin{deluxetable}{cccc}
\tablecaption{Kinematic properties of the model galaxies.
\label{dyn}
}
\tablehead{
\colhead{$C$$^{\rm a}$} & \colhead{$(\sigma_z/\sigma_x)$$^{\rm b}$} &
\colhead{spin ($\lambda$)} & \colhead{stellar mass $^{\rm c}$}
}
\startdata
0.1 & 0.59 & 0.70 & $4.2 \times 10^{10}$ \SM\\
0.5 & 0.71 & 0.52 & $5.4 \times 10^{10}$ \SM\\
1.0 & 0.77 & 0.44 & $4.6 \times 10^{10}$ \SM\\
\enddata
\tablecomments{
$^{\rm a}$ The star formation parameter.\\
$^{\rm b}$ The ratio between the X and Z axis velocity dispersion.\\
$^{\rm c}$ The stellar mass at $t = 5$ Gyr.\\
}
\end{deluxetable}

\clearpage

\clearpage

\end{document}